\documentclass[sigplan,screen,authorversion,nonacm]{acmart}

\setcopyright{acmcopyright}
\copyrightyear{2018}
\acmYear{2018}
\acmDOI{XXXXXXX.XXXXXXX}
\acmConference[]{}{}{}
\usepackage{multirow}
\usepackage{amsmath}
\usepackage{bm}

\begin{document}

\title{Accurate battery lifetime prediction across diverse aging conditions with deep learning}
\author{Han Zhang$^{1\star}$, Yuqi Li$^{2\star}$, Shun Zheng$^3$, Ziheng Lu$^3$, Xiaofan Gui$^3$, Wei Xu$^1$, Jiang Bian$^3$}
\affiliation{
\institution{$^1$Institute for Interdisciplinary Information Sciences, Tsinghua University}
\institution{$^2$Department of Materials Science and Engineering, Stanford University}
\institution{$^3$Microsoft Research}
\country{}
}

\renewcommand{\shortauthors}{Zhang and Li, et al.}
\newcommand{\todo}[1]{\textcolor{red}{#1}}
\newcommand{\zhanghan}[1]{\textcolor{blue}{#1}}
\newcommand{\zheng}[1]{\textcolor{black}{#1}}
\newcommand{\bian}[1]{\textcolor{black}{#1}}
\newcommand{\note}[1]{\textcolor{red}{[NOTE: #1]}}

\newcommand\myauthornote[1]{%
  \begingroup
  \renewcommand\thefootnote{}\footnote{#1}%
  \addtocounter{footnote}{-1}%
  \endgroup
}

\newcommand{\methodname}{\texttt{BatLiNet}}

\begin{abstract}
Accurately predicting the lifetime of battery cells in early cycles holds tremendous value for battery research and development as well as numerous downstream applications~\cite{NE,attia2021statistical,CLO,Yao2023NRM}.
This task is rather challenging because diverse conditions, such as electrode materials, operating conditions, and working environments, collectively determine complex capacity-degradation behaviors.
However, current prediction methods are developed and validated under limited aging conditions~\cite{NE,attia2021statistical,HUST}, resulting in questionable adaptability to varied aging conditions and an inability to fully benefit from historical data collected under different conditions.
Here we introduce a universal deep learning approach that is capable of accommodating various aging conditions and facilitating effective learning under low-resource conditions by leveraging data from rich conditions.
Our key finding is that incorporating inter-cell feature differences, rather than solely considering single-cell characteristics, significantly increases the accuracy of battery lifetime prediction and its cross-condition robustness.
Accordingly, we develop a holistic learning framework accommodating both single-cell and inter-cell modeling.
A comprehensive benchmark is built for evaluation, encompassing 401 battery cells utilizing 5 prevalent electrode materials across 168 cycling conditions.
We demonstrate remarkable capabilities in learning across diverse aging conditions, exclusively achieving 10\% prediction error using the first 100 cycles, and in facilitating low-resource learning, almost halving the error of single-cell modeling in many cases.
More broadly, by breaking the learning boundaries among different aging conditions, our approach could significantly accelerate the development and optimization of lithium-ion batteries.
\end{abstract}




\maketitle

\myauthornote{$^\star$ Equal contribution. This work was done during the internships of Han Zhang and Yuqi Li at Microsoft Research.}

Owing to their high energy densities and low production costs, lithium-ion batteries have been widely adopted in modern industry, propelling the surge of renewable energy solutions and electric vehicles\cite{Tarascon2001IssuesAC,Armand2008BuildingBB,Dunn2011ElectricalES}.
Nevertheless, the capacity of lithium-ion batteries inevitably fades with cyclic operations due to their intrinsic electrochemical mechanisms.
Unexpected rapid degradation not only leads to poor user experiences, such as range anxiety for electric vehicles, but can also affect the operation of essential facilities, such as the stability of power grids.
To proactively mitigate these side effects, accurately predicting battery lifetime in early cycles has been identified as a critical task\cite{ng2020soc_soh_nmi,Yao2023NRM}.
This task is rather challenging because numerous factors, including electrode materials, charging and discharging protocols, and working environments, collectively influence the complex battery aging process.
Recent data-driven approaches that leverage machine learning have made remarkable progress in this direction~\cite{NE,attia2021statistical,CLO,HUST}, identifying critical electrical features that highly correlate with cycle life.

However, existing methods for battery lifetime prediction have been developed and validated under limited aging conditions, such as testing only lithium-iron-phosphate ($LiFePO_4$) batteries and using single charging or discharging protocols~\cite{NE,attia2021statistical,HUST}.
Data characteristics under these restricted conditions affect feature extraction and model design, potentially limiting the success and generalization of their conclusions.
It remains questionable whether these methods perform well under varied aging conditions.
Moreover, focusing on limited aging conditions restricts the research development of leveraging historical data collected under different conditions.
This limitation separates battery datasets emphasizing different aging factors as isolated islands, hindering the development of general modeling approaches.


Here we introduce a deep learning framework, \methodname, tailored to predict battery lifetime across a variety of aging conditions—including electrode materials, cycling protocols, and temperature fluctuations. At its core, the framework innovates with "inter-cell learning," which contrasts pairs of batteries to discern lifetime differences, a significant leap from traditional models that focus solely on individual "intra-cell" data. This dual approach not only captures individual degradation patterns but also contextualizes them within a broader, comparative aging landscape.

Our findings demonstrate that this inter-cell perspective crucially enhances the model's predictive precision and robustness, especially under conditions where data is sparse, such as with novel electrode materials. By integrating the comparative inter-cell strategy with the conventional intra-cell analysis into a singular, unified framework, we bridge the gap between isolated and relative aging scenarios.

The neural networks embedded within \methodname~adeptly navigate the nonlinear intricacies inherent to battery aging, ensuring the framework's adeptness in learning from limited resources. This adaptability, coupled with our method's attention to diverse aging conditions, positions \methodname~as a comprehensive solution for accurate, resilient battery lifetime predictions, essential for advancing reliable energy storage systems.

To validate the effectiveness of \methodname, we construct a comprehensive benchmark by collecting all public datasets~\cite{NE,CLO,SNL,HUST,CALCE1,CALCE2,CALCE3} that emphasize different aging conditions and contain the necessary information to support learning and evaluation.
To the best of our knowledge, this benchmark is the largest and most diverse in terms of aging conditions for battery lifetime prediction.
Our results demonstrate the remarkable effectiveness of \methodname~in robustly producing accurate predictions across diverse aging conditions and in boosting prediction performance for low-resource conditions.
Regarding learning across diverse aging conditions, our approach exclusively achieves a 10\% prediction error using the first 100 cycles.
Moreover, as for learning in low-resource scenarios, we use $2$ to $8$ cells to achieve $20\%$ prediction error in predicting the lifetime of new batteries, halving the error of direct learning on rare data in many cases.
\section*{The BatLiNet Framework}

\begin{figure*}[!t]
\includegraphics[width=\linewidth]{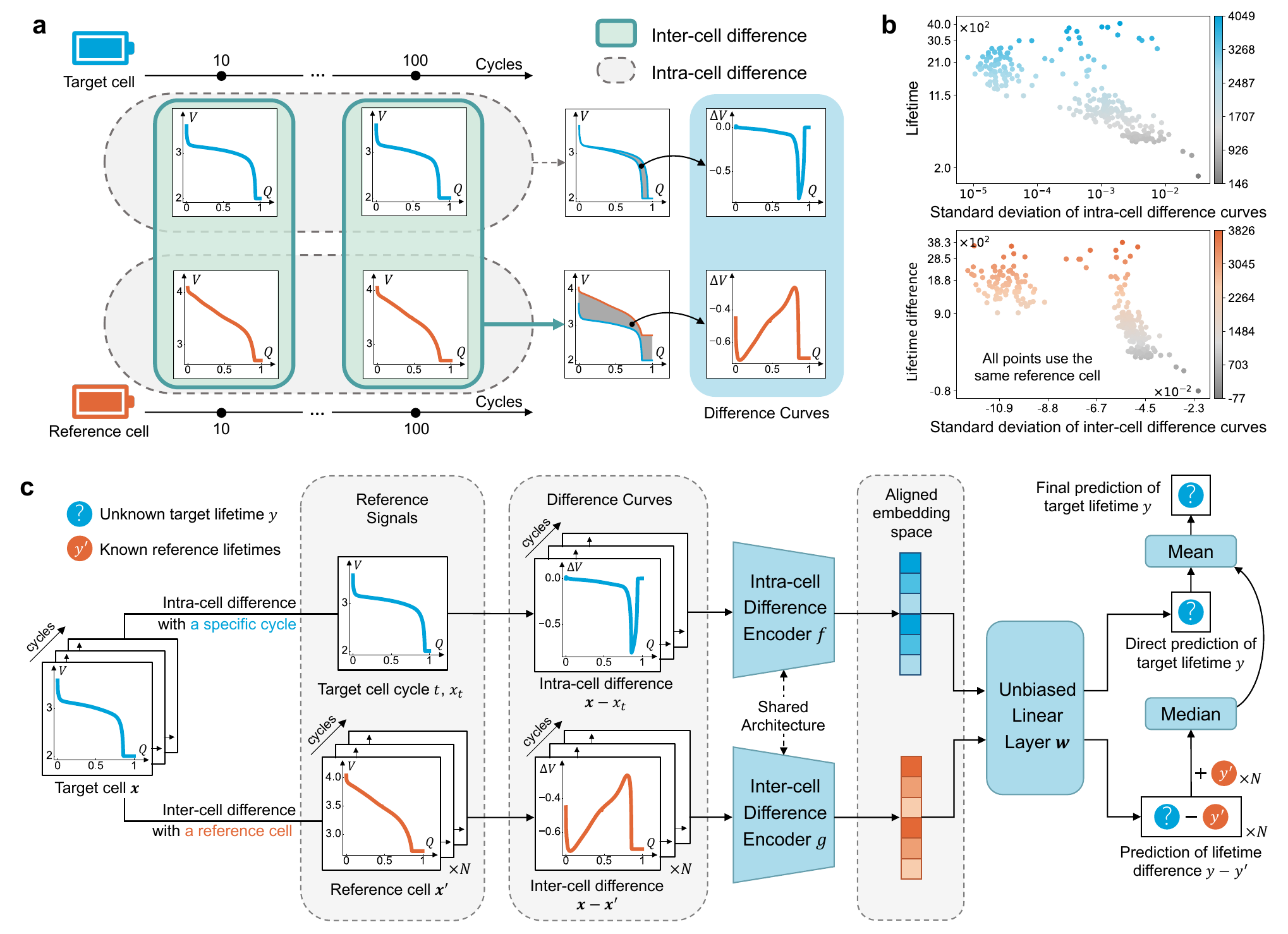}
\caption{An overview of the BatLiNet framework, where we adopt a lithium-iron-phosphate battery cell as the target cell and leverage another lithium-cobalt-oxide battery cell as the reference. \textbf{a}: The feature construction for intra-cell and inter-cell learning. \textbf{b}: The correlations between constructed features and prediction labels for both intra-cell (the upper part) and inter-cell (the lower part) learning. \textbf{c}: The overall pipeline of the BatLiNet framework.}
\label{fig:pipeline}
\end{figure*}

BatLiNet is a holistic framework that unifies intra-cell and inter-cell learning.
Intra-cell learning involves capturing critical early-cycle patterns of a single battery cell that highly indicate its lifetime, while inter-cell learning aims to characterize the differences between a pair of battery cells in early cycles, implying their lifetime difference.
Though serving different objectives, inter-cell learning can be directly leveraged for battery lifetime prediction by treating one cell as a target and regarding the other as a reference, which is assumed to have a known lifetime.

In \figurename~\ref{fig:pipeline}, we provide an overview of the BatLiNet framework.
Specifically, \figurename~\ref{fig:pipeline}a depicts the process of constructing effective features for intra-cell and inter-cell learning, where we use the voltage-capacity curve during the discharge stage as a typical example to represent cycle-level features.
Traditional approaches for intra-cell learning operate on a single cell and obtain effective intra-cell features by calculating the cycle-level feature differences between an early cycle, such as the 100-th cycle, and an initial cycle, such as the 10-th cycle~\cite{NE,attia2021statistical}.
In this work, we refer to these operations as calculating \emph{intra-cell differences}, meaning the feature differences between different cycles within a single cell.
To adapt to diverse aging conditions, we propose to calculate \emph{inter-cell differences} by contrasting the cycle-level features between a target cell and a reference cell in the same cycle.
The inter-cell differences help to enhance the understanding of the relations and differences between different conditions.
Besides, in addition to the discharging voltage-capacity curve, other cycle-level feature maps, such as the charging voltage-capacity curve, can also be utilized in both intra-cell and inter-cell learning (detailed in Methods).

To understand the effectiveness of intra-cell and inter-cell differences across diverse aging conditions, we visualize the connections between these difference curves and their corresponding targets for all battery cells in \figurename~\ref{fig:pipeline}b.
In the upper part of \figurename~\ref{fig:pipeline}b, we plot the correlation between the standard deviation of intra-cell difference curves and the corresponding lifetime.
Similar to the observations in previous studies~\cite{NE,attia2021statistical}, the standard deviation of intra-cell difference curves for a specific cell roughly presents a near-linear correlation with its lifetime.
However, we additionally observe more scattered points and more outliers due to the incorporation of much more diverse aging conditions.
These new observations intuitively indicate the inadaptability of past experiences to different aging conditions.
Moreover, we plot the connection between inter-cell differences and associated lifetime differences in the lower part of \figurename~\ref{fig:pipeline}b, where all inter-cell differences use the same reference cell.
Similarly, we also observe some near-linear components, non-linear distributions, and outliers, which call for sophisticated modeling beyond linear approaches.
More importantly, we can find that the relationship between inter-cell differences and lifetime differences presents a very different distribution than the corresponding intra-cell relationship, indicating that learning in intra-cell and inter-cell spaces can be complementary to each other.

Accordingly, we design a two-branch encoding architecture followed by a shared linear layer to unify intra-cell and inter-cell learning, as shown in \figurename~\ref{fig:pipeline}c.
Given a target cell, we feed its cycle-level features into two separate branches.
The first branch calculates intra-cell difference curves by comparing features in all early cycles with that of a reference cycle and employs a specific neural network $f$ for intra-cell learning.
In addition, the second branch includes a reference cell to obtain inter-cell difference curves and further feeds them into another neural network $g$ for inter-cell learning.
After the encoding procedures of $f$ and $g$, the resulting representations are mapped to an aligned linear space, allowing us to leverage a shared linear layer to emit outputs for these two branches (we provide the derivation for the necessity of this shared linear layer in Methods).
During training, we can randomly assign reference cells to stimulate robust inter-cell modeling.
At inference time, we can randomly select references from training cells and use the average of outputs from these two branches as the final prediction.
With these customized designs, the \methodname framework adapts to diverse aging conditions and robustly produce accurate battery lifetime predictions.

\section*{A Comprehensive Benchmark Covering Diverse Aging Conditions}

\begin{figure*}[!t]
    \includegraphics[width=\linewidth]{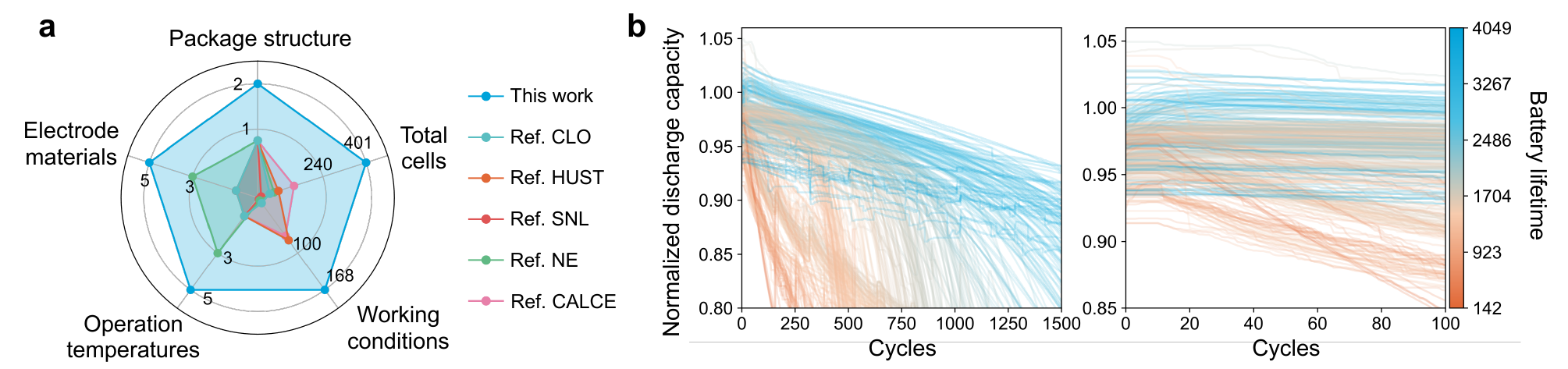}
    \caption{Visualization of datasets. \textbf{a}: The coverage of the dataset employed in this work significantly surpasses that of previous undertakings, namely the NE\cite{NE}, CLO\cite{CLO}, SNL\cite{SNL}, HUST\cite{HUST}, and CALCE\cite{CALCE1,CALCE2,CALCE3}. \textbf{b}: A visualization of the capacity degradation with respect to the cycles for all the cells. The curves are normalized by their nominal capacities. The first $1500$ and $100$ cycles are shown. The degradation patterns are highly non-linear and are difficult to distinguish in the first $100$ cycles.}
    \label{fig:dataset}
\end{figure*}

To comprehensively evaluate and compare \methodname against existing modeling approaches, we curated, to the best of our knowledge, all publicly available battery data for battery lifetime prediction into five benchmark datasets. 

\texttt{MATR-1}, \texttt{MATR-2}, and \texttt{HUST} consist of commercial 18650 lithium-ion phosphate/graphite (LFP) batteries of the same model. \texttt{MATR-1} and \texttt{MATR-2} use the same training set and evaluate the prediction performance on trained or unseen charging protocols, respectively.
The batteries in \texttt{HUST} employ an identical charging protocol but different discharge rates to examine the model's generalization capability across diverse discharge protocols.

Additionally, we collected battery data used in prior studies including CLO~\cite{CLO}, CALCE~\cite{CALCE1,CALCE2,CALCE3}, HNEI~\cite{HNEI}, UL-PUR~\cite{ULPUR}, RWTH~\cite{RWTH} and SNL~\cite{SNL}.
By combining these batteries with \texttt{MATR-1}, \texttt{MATR-2} and \texttt{HUST}, we obtained a total collection of 401 batteries and developed two datasets \texttt{MIX-100} and \texttt{MIX-20}.
\texttt{MIX-100} examines the typical early prediction case where models must forecast the $80\%$ end-of-life point using just the first $100$ cycles of data.
\texttt{MIX-20} poses a more difficult challenge - models must predict the number of cycles before capacity degrades to $90\%$ of nominal, using only the first $20$ cycles.
Notably, any batteries reaching end-of-life prematurely during the initial cycles were excluded from the experiments. 

\figurename~\ref{fig:dataset}\textbf{a} shows that the collected dataset exhibits a more comprehensive coverage that goes beyond the confines of previous works.
\figurename~\ref{fig:dataset}\textbf{b} further visualizes the capacity degradation of the collected batteries in the first $1500$ and $100$ cycles of the batteries.
Here the long-term degradation is highly non-linear and demonstrates considerable variation among batteries.
In contrast, the degradation within the initial $100$ cycles remains indiscernible for most batteries.
Such diverse and intricate characteristics require the model to capture and distinguish the degradation patterns caused by various factors for generalization.

\section*{Accurate Battery Lifetime Prediction Empowered by \methodname}

\begin{table*}[!t]
\small
\begin{tabular}{lllllllllll}
\toprule
{\multirow{2}{*}{Method}} & \multicolumn{2}{c}{MATR-1} & \multicolumn{2}{c}{MATR-2}  & \multicolumn{2}{c}{HUST} & \multicolumn{2}{c}{MIX-100} & \multicolumn{2}{c}{MIX-20} \\
{} &    RMSE & MAPE(\%) &     RMSE & MAPE(\%) &      RMSE & MAPE(\%) &     RMSE & MAPE(\%) &    RMSE & MAPE(\%) \\
\midrule
Training Mean                 &     399 &      28 &      511 &      36 &      420 &      18 &      573 &     59 &     593 &      102 \\
\midrule
``Variance'' Model\cite{NE}        &     138 &      15 &      196 &      12 &        398 &      17 &      521 &      39 &     601 &      95 \\
``Discharge'' Model\cite{NE}       &      \underline{86} &       \underline{8} &      \underline{173} &      \underline{11} &       \underline{322} &      \underline{14} &  1743 &    47 &    >2000 &      >100 \\
``Full'' Model\cite{NE}            &     100 &      11 &      214 &      12 &         335 &      14 &      331 &      22 &     441 &      53 \\
Ridge Regression\cite{attia2021statistical}      &     125 &      13 &      188 &      11 &       1047 &      36 &      395 &     30 &     806 &      150 \\
PCR\cite{attia2021statistical}                   &     100 &      11 &      176 &      11 &      435 &      19 &      384 &      28 &     701 &      78 \\
PLSR\cite{attia2021statistical}                  &      97 &      10 &      193 &      11 &      431 &      18 &      371 &      26 &     543 &      77 \\
SVM Regression        &     140 &      15 &      300 &      18 &     344 &      16 &      257 &      18 &     438 &      46 \\
Random Forest\cite{attia2021statistical}         &     140 &      15 &      202 &      11 &      348 &      16 &      \underline{211} &      \underline{14} &     \underline{288} &      \underline{31} \\
\midrule
MLP\cite{attia2021statistical}                   &   162±7 &    12±0 &    207±4 &    11±0 &   444±5 &    18±1 &  455±37 &    27±1 &   532±25 &    61±6\\
LSTM                  &  123±11 &    12±2 &   226±36 &    14±2 &    442±32 &    20±1 &   266±11 &    15±1 & 417±62 &    37±7 \\
CNN\cite{attia2021statistical}                   &  115±96 &     9±6 &  237±107 &    17±8 &    445±35 &    21±1  &  261±38 &    15±1 &  785±132 &    41±4 \\
\midrule
\texttt{BatLiNet} (ours) &    \textbf{59±2} &     \textbf{6±0} &   \textbf{163±12} &    \textbf{11±1} &    \textbf{264±9} &    \textbf{10±1} &   \textbf{158±7} &    \textbf{10±0} &   \textbf{201±18} &    \textbf{18±1}\\
\bottomrule
\end{tabular}
\caption{Comparison of prediction errors with baselines. We employ bold font to emphasize the best-performing method and utilize underlines to denote the second-best methods. For neural-network-based methods, we report the mean and standard deviation of eight random seeds.}
\label{tab:main}
\end{table*}

\begin{figure*}
    \includegraphics[width=0.8\linewidth]{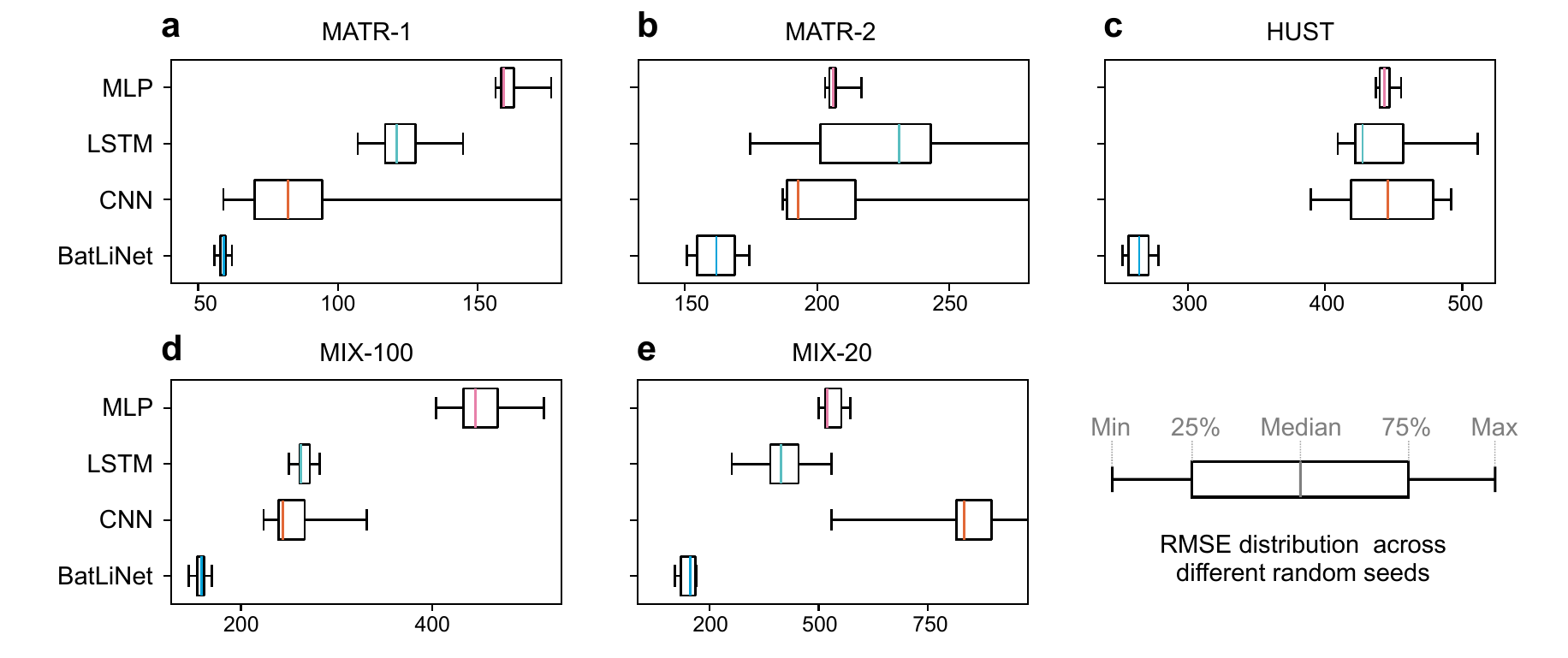}
    \caption{Distribution of prediction error for deep learning models across five benchmark datasets. Box plots illustrate the distribution of root mean squared error (RMSE) over eight random initializations of each deep architecture.}
    \label{fig:error_dist}
\end{figure*}

In Table~\ref{tab:main}, we present the performance comparison of \methodname \ against various baselines on the five benchmark datasets.
The ``variance'', ``discharge'', and ``full'' models~\cite{NE} employ linear regression on statistical features derived from discharge capacity-voltage curves.
Ridge Regression instead directly fits a linear model to the raw discharge curves~\cite{attia2021statistical}.
Partial Least-Squares Regression (PLSR) \cite{Geladi1986PartialLR} and Principal Component Regression (PCR) 
\cite{Shen2009PrincipalCA} first group cells based on their raw discharge curves and then fit linear models on the result groups.
The Support Vector Machine (SVM) \cite{Drucker1996SupportVR} and Random Forest \cite{Breiman2001RandomF} are non-linear statistical models with a stronger fitting ability.
For deep models, we compare \methodname against Multi-layer Perceptron (MLP)\cite{attia2021statistical}, the Long-Short Term Memory network (LSTM)\cite{Hochreiter1997LongSM}, and Convolutional Networks (CNN)\cite{Krizhevsky2012ImageNetCW}.

The "Discharge" model demonstrates strong performance on \texttt{MATR-1}, \texttt{MATR-2}, and \texttt{HUST}, yet struggles with diverse chemistries and intricate aging patterns in \texttt{MIX-100} and \texttt{MIX-20}.
In comparison, Ridge Regression shows lower prediction error when applied across battery types, indicating that such LFP-focused features may not effectively generalize to diverse aging conditions.
By explicitly differentiating cells beforehand, PLSR and PCR incorporate nonlinear complexity into linear frameworks and further improve over Ridge Regression.
Both SVM and RF excel on \texttt{MIX-100} and \texttt{MIX-20} compared to other linear baselines, confirming that the complex degradation patterns call for models with stronger fitting capability.

The deep learning models exhibit high variability in performance depending on random initialization.
As shown in \figurename~\ref{fig:error_dist}, the error distribution of the deep models fluctuates widely across random seeds.
The MLP architecture demonstrates relatively modest prediction errors with the lowest variance on all datasets.
In contrast, CNN displays the largest variance in performance, yet achieves the lowest errors on three datasets among baselines given optimal seeds~\cite{attia2021statistical}.
LSTM strikes a favorable balance between accuracy and variance.
However, all deep models are susceptible to overfitting and exhibit minimal predictive advantage compared to conventional statistical approaches regarding both regression accuracy and error variance.

While no single technique consistently optimizes performance across datasets, \methodname achieves $\leq 11\%$ mean absolute percentage error (MAPE) given the first $100$ cycles and $\leq 18\%$ for the first $20$ cycles.
This reduces the root mean square error (RMSE) versus the top baseline by $31.4\%$, $5.8\%$, $18.0\%$, $30.2\%$, and $25.11\%$ on the five datasets, respectively. 
By jointly training the intra-cell and inter-cell branches, \methodname combines strong fitting capabilities beyond conventional statistical models with robustness to random initialization lacking in existing deep models.
Through this blend of strengths, \methodname delivers accurate and reliable lifetime predictions across diverse battery aging conditions.

\section*{Facilitating Effective Learning in Low-resource Scenarios}

\begin{figure*}
    \includegraphics[width=\linewidth]{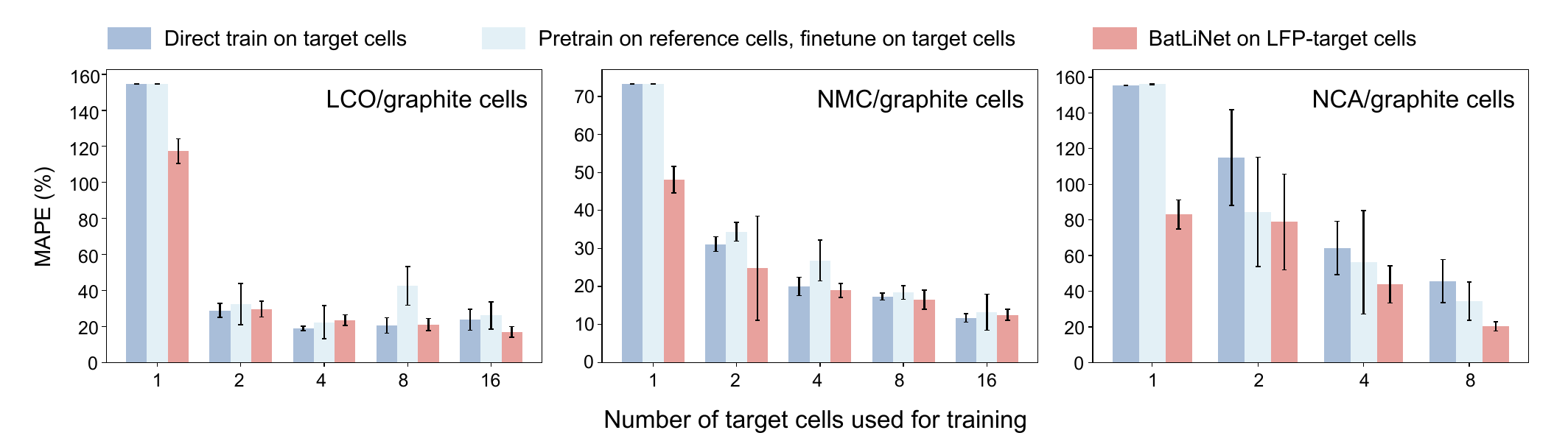}
    \caption{Predictive performance of three learning paradigms that employ LFP/graphite cells to benefit learning in low-resource scenarios. We report the mean and variance of the percentage error under eight random seeds.}
    \label{fig:low_resource}
\end{figure*}

In practice, battery development is often constrained by limited resources, resulting in a restricted number of available test batteries.
To develop an accurate life prediction model, cycling a considerable number of batteries to end-of-life is required for label collection.
However, this process is both time-consuming and unsustainable.
An alternative methodology is to leverage the abundant historical data of batteries with varying aging states to augment model training under such low-resource scenarios.

To simulate resource-constrained applications, the cells in \texttt{MIX-100} were sorted by cathode material, and the 275 LFP/graphite cells were leveraged to augment the prediction performance on the $37$ LCO cells, $22$ NCA cells, and $69$ NMC cells.
Among the test batteries, $21$ LCO cells, $14$ NCA cells, and $53$ NMC cells were randomly sampled for model evaluation, and we employ $1$, $2$, $4$, $8$, and $16$ cells from the remainder to simulate varying cycling test budgets.
Three learning paradigms were investigated to comprehensively analyze the influence of historical data: 1) direct training of a CNN on the target cells, 2) transfer learning where a CNN was pre-trained on LFP cells then fine-tuned on the target cells, and 3) training \methodname on the combined LFP and target cells.


\figurename~\ref{fig:low_resource} presents the performance of the three learning paradigms under resource-constrained conditions.
Pre-training the model on historical LFP/graphite cell data substantially reduced prediction error by $10\%$ for NCA cells.
However, it resulted in inferior predictions for LCO and NMC cells, suggesting that degradation patterns vary between battery chemistries.
In contrast, \methodname leveraged the LFP cells to improve over direct learning for all three cathode materials, indicating the inter-cell modeling between LFP-target pairs generalizes across cathode materials.
Notably, \methodname further reduced prediction error by $10\%$ over transfer learning, achieving $20.26\%$ MAPE using only two to eight target cells and historical LFP data.
This high data efficiency greatly mitigates the data hunger in battery development and can substantially reduce downstream application modeling costs such as developing novel electrode materials or electrolytes.
In contrast, Attia et al. \cite{CLO} developed a linear model trained on a dataset of $41$ LFP cells when optimizing the rapid charging protocol specifically for batteries possessing the same lithium iron phosphate cathode chemistry.

\section*{Working Mechanisms of \methodname}

\begin{figure*}[!t]
    \includegraphics[width=\linewidth]{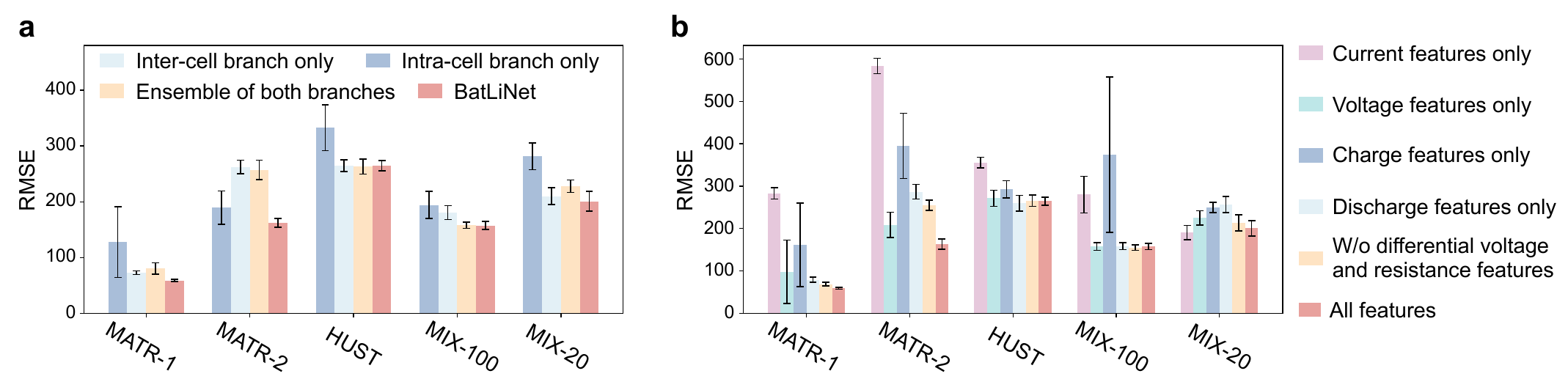}
    \caption{Ablation studies to illustrate the mechanism of \methodname.
    \textbf{a}: Leveraging the joint training of intra-cell and inter-cell branches, \methodname achieves superior performance than the ensemble of separately trained inter-cell and intra-cell branches.
    \textbf{b}: Enhanced predictive performance necessiates the inclusion of the complete voltage and current signals, as well as their interactions, from both charging and discharging stages.
    }
    \label{fig:ablation}
\end{figure*}

Here we conduct further ablation experiments to study the mechanism of \methodname.

\figurename~\ref{fig:ablation}\textbf{a} compares the predictive performance of different modeling branches in \methodname.
Across all datasets, intra-cell models exhibit the highest error and variance.
The inter-cell model shows good robustness with lower variance across all datasets, yet its prediction accuracy is slightly lower than that of the intra-cell modeling on some datasets, as its predictions are implicitly derived through the reference cells.
The ensemble of the two separately trained branches obtains better performance on \texttt{HUST} and \texttt{MIX-100}, yet on other datasets it inherits the flaws of either branch, resulting in higher variance or suboptimal accuracy.
\methodname achieved the most robust and precise lifetime estimation across all datasets, demonstrating that through the co-training with shared prediction layer, the model better integrates the strength of both branches and yields the strongest predictive performance.

\figurename~\ref{fig:ablation}\textbf{b} displays the performance of \methodname when using a subset of the six capacity-indexed features.
It's evident that all six features play a crucial role in achieving improved predictive performance. 
The current-based features can lead to high variance and large errors, yet they are indispensable for the extreme early prediction scenarios in \texttt{MIX-20}.
The voltage- and discharge-based features demonstrate strong performance across all datasets.
Merging voltage and current features from both charge and discharge stages results in enhanced performance, with the best outcomes achieved by employing all six features.

\section*{Conclusion}

By introducing inter-cell learning and unifying it with intra-cell learning, our framework \methodname~has significantly boosted the performance of battery lifetime prediction across diverse aging conditions.
It is noteworthy that the proposed inter-cell learning, enabling the modeling of the relations and differences among diverse aging conditions, not only provides exclusive yet complementary value for traditional intra-cell learning but also facilitates the effective knowledge transfer from rich conditions with abundant historical data to low-resource scenarios of emerging needs.
Looking into the future, our method can be leveraged in various developing aspects that lead to varied aging conditions, such as different fast charging protocols~\cite{zeng2023extreme_fast_charging} and new electrode materials~\cite{shraer2022dev_pos_material}.
Specifically, as an accurate and robust lifetime predictor across different aging conditions, our method holds the largely improved capability in accelerating the development and optimization of lithium-ion batteries~\cite{CLO}.
Moreover, the idea of modeling inter-cell differences can be extended to other crucial prediction tasks, such as predicting the state of charge and health~\cite{ng2020soc_soh_nmi,Roman2021NMI_soh_pipeline}, broadly benefiting battery management and applications.

\bibliography{main}

\begin{thebibliography}{10}

\bibitem{NE}
K.~A. Severson, P.~M. Attia, N.~Jin, N.~Perkins, B.~Jiang, Z.~Yang, M.~H. Chen,
  M.~Aykol, P.~K. Herring, D.~Fraggedakis, M.~Z. Bazant, S.~J. Harris, W.~C.
  Chueh, and R.~D. Braatz, ``Data-driven prediction of battery cycle life
  before capacity degradation,'' {\em Nature Energy}, vol.~4, pp.~383--391,
  2019.

\bibitem{attia2021statistical}
P.~M. Attia, K.~A. Severson, and J.~D. Witmer, ``Statistical learning for
  accurate and interpretable battery lifetime prediction,'' {\em Journal of The
  Electrochemical Society}, vol.~168, no.~9, p.~090547, 2021.

\bibitem{CLO}
P.~M. Attia, A.~Grover, N.~Jin, K.~A. Severson, T.~M. Markov, Y.-H. Liao, M.~H.
  Chen, B.~Cheong, N.~Perkins, Z.~Yang, P.~K. Herring, M.~Aykol, S.~J. Harris,
  R.~D. Braatz, S.~Ermon, and W.~C. Chueh, ``Closed-loop optimization of
  fast-charging protocols for batteries with machine learning,'' {\em Nature},
  vol.~578, pp.~397--402, 2020.

\bibitem{Yao2023NRM}
Z.~Yao, Y.~Lum, A.~Johnston, L.~M. Mejia-Mendoza, X.~Zhou, Y.~Wen,
  A.~Aspuru-Guzik, E.~H. Sargent, and Z.~W. Seh, ``Machine learning for a
  sustainable energy future,'' {\em Nature Reviews Materials}, vol.~8, no.~3,
  pp.~202--215, 2023.

\bibitem{HUST}
G.~Ma, S.~Xu, B.~Jiang, C.~Cheng, X.~Yang, Y.~Shen, T.~Yang, Y.~Huang, H.~Ding,
  and Y.~Yuan, ``Real-time personalized health status prediction of lithium-ion
  batteries using deep transfer learning,'' {\em Energy \& Environmental
  Science}, 2022.

\bibitem{Tarascon2001IssuesAC}
J.~M. Tarascon and M.~Armand, ``Issues and challenges facing rechargeable
  lithium batteries,'' {\em Nature}, vol.~414, pp.~359--367, 2001.

\bibitem{Armand2008BuildingBB}
M.~Armand and J.~M. Tarascon, ``Building better batteries,'' {\em Nature},
  vol.~451, pp.~652--657, 2008.

\bibitem{Dunn2011ElectricalES}
B.~S. Dunn, H.~Kamath, and J.~M. Tarascon, ``Electrical energy storage for the
  grid: A battery of choices,'' {\em Science}, vol.~334, pp.~928 -- 935, 2011.

\bibitem{ng2020soc_soh_nmi}
M.-F. Ng, J.~Zhao, Q.~Yan, G.~J. Conduit, and Z.~W. Seh, ``Predicting the state
  of charge and health of batteries using data-driven machine learning,'' {\em
  Nature Machine Intelligence}, vol.~2, no.~3, pp.~161--170, 2020.

\bibitem{SNL}
Y.~Preger, H.~M. Barkholtz, A.~Fresquez, D.~L. Campbell, B.~W. Juba, J.~K.
  Rom{\'a}n-Kustas, S.~R. Ferreira, and B.~R. Chalamala, ``Degradation of
  commercial lithium-ion cells as a function of chemistry and cycling
  conditions,'' {\em Journal of The Electrochemical Society}, vol.~167,
  p.~120532, 2020.

\bibitem{CALCE1}
W.~He, N.~Williard, M.~Osterman, and M.~Pecht, ``Prognostics of lithium-ion
  batteries based on dempster–shafer theory and the bayesian monte carlo
  method,'' vol.~196(23), pp.~10314--10321, 2011.

\bibitem{CALCE2}
Y.~Xing, E.~Ma, K.~L. Tsui, and M.~Pecht, ``An ensemble model for predicting
  the remaining useful performance of lithium-ion batteries,'' vol.~53(6),
  pp.~811--820, 2013.

\bibitem{CALCE3}
N.~Williard, W.~He, M.~Osterman, , and M.~Pecht, ``Comparative analysis of
  features for determining state of health in lithium-ion batteries,'' vol.~4,
  pp.~1--7, 2013.

\bibitem{HNEI}
A.~Devie, G.~Baure, and M.~Dubarry, ``Intrinsic variability in the degradation
  of a batch of commercial 18650 lithium-ion cells,'' {\em Energies}, vol.~11,
  p.~1031, 2018.

\bibitem{ULPUR}
D.~Juarez-Robles, J.~A. Jeevarajan, and P.~P. Mukherjee, ``Degradation-safety
  analytics in lithium-ion cells: Part i. aging under charge/discharge
  cycling,'' {\em Journal of The Electrochemical Society}, vol.~167, p.~160510,
  2020.

\bibitem{RWTH}
W.~Li, N.~Sengupta, P.~A. Dechent, D.~Howey, A.~Annaswamy, and D.~U. Sauer,
  ``{O}ne-shot battery degradation trajectory prediction with deep learning,''
  {\em Journal of power sources}, p.~230024, 2021.

\bibitem{Geladi1986PartialLR}
P.~Geladi and B.~R. Kowalski, ``Partial least-squares regression: a tutorial,''
  {\em Analytica Chimica Acta}, vol.~185, pp.~1--17, 1986.

\bibitem{Shen2009PrincipalCA}
H.~T. Shen, ``Principal component analysis,'' in {\em Encyclopedia of Database
  Systems}, 2009.

\bibitem{Drucker1996SupportVR}
H.~Drucker, C.~J.~C. Burges, L.~Kaufman, A.~Smola, and V.~N. Vapnik, ``Support
  vector regression machines,'' in {\em Neural Information Processing Systems},
  1996.

\bibitem{Breiman2001RandomF}
L.~Breiman, ``Random forests,'' {\em Machine Learning}, vol.~45, pp.~5--32,
  2001.

\bibitem{Hochreiter1997LongSM}
S.~Hochreiter and J.~Schmidhuber, ``Long short-term memory,'' {\em Neural
  Computation}, vol.~9, pp.~1735--1780, 1997.

\bibitem{Krizhevsky2012ImageNetCW}
A.~Krizhevsky, I.~Sutskever, and G.~E. Hinton, ``Imagenet classification with
  deep convolutional neural networks,'' {\em Communications of the ACM},
  vol.~60, pp.~84 -- 90, 2012.

\bibitem{zeng2023extreme_fast_charging}
Y.~Zeng, B.~Zhang, Y.~Fu, F.~Shen, Q.~Zheng, D.~Chalise, R.~Miao, S.~Kaur,
  S.~D. Lubner, M.~C. Tucker, {\em et~al.}, ``Extreme fast charging of
  commercial li-ion batteries via combined thermal switching and self-heating
  approaches,'' {\em Nature Communications}, vol.~14, no.~1, p.~3229, 2023.

\bibitem{shraer2022dev_pos_material}
S.~D. Shraer, N.~D. Luchinin, I.~A. Trussov, D.~A. Aksyonov, A.~V. Morozov,
  S.~V. Ryazantsev, A.~R. Iarchuk, P.~A. Morozova, V.~A. Nikitina, K.~J.
  Stevenson, {\em et~al.}, ``Development of vanadium-based polyanion positive
  electrode active materials for high-voltage sodium-based batteries,'' {\em
  Nature Communications}, vol.~13, no.~1, p.~4097, 2022.

\bibitem{Roman2021NMI_soh_pipeline}
D.~Roman, S.~Saxena, V.~Robu, M.~Pecht, and D.~Flynn, ``Machine learning
  pipeline for battery state-of-health estimation,'' {\em Nature Machine
  Intelligence}, vol.~3, no.~5, pp.~447--456, 2021.

\end{thebibliography}
\bibliographystyle{ieeetr}

\newpage
\section*{Methods}

\subsection*{Building Cycle-level Feature Maps}

We have shown how to build features for intra-cell and inter-cell learning in \figurename~\ref{fig:pipeline}a, by leveraging the discharging voltage-capacity curve as an example for cycle-level features.
Here we include a comprehensive discussion about the procedure of building cycle-level feature maps.

The raw data available for each cycle includes various electrical signals collected during the charging and discharging process.
However, there are several obstacles preventing us from directly utilizing these raw signals.
First, the electrical signals in different cycles usually have variable lengths. Hence, it is necessary to align them properly into a regular format to fit for neural networks.
Meanwhile, due to the fine-grained recording of electrical signals, the raw high-dimensional features are highly redundant and require effective noise reduction.
Besides, each cycle contains various electrical signals (voltage, current, states of charge/discharge) on two different stages (charge, discharge), which call for effective representations that cannot only contain valuable electrical signals but also yield the ability to differentiate respective values of various signal types within the same or between different stages.

With full consideration of these obstacles, we develop specific processing steps to obtain cycle-level feature maps from the raw electrical signals.
The core idea is to use the normalized state of charge ($Q$) as a new index to align different electrical signals in different cycles and compute their differentiation within the same stage or between charge and discharge stages.
In this way, we do not need to worry about the variable lengths across cycles.
Besides, due to the redundant recording of electrical signals, we can obtain these $Q$-indexed series by performing interpolations on raw time-indexed signals and easily control the feature dimensions by adjusting interpolation granularity.
After this step, we can obtain four types of processed series indexed by $Q$, $V_c(Q)$, $V_d(Q)$, $I_c(Q)$, and $I_d(Q)$, where $Q$ varies from 0 to 1 with a pre-specified step.
We further derive two additional signals to characterize the connections between the charge and discharge stages:
1) $\Delta V(Q) = V_c(Q) - V_d(Q)$, meaning the gap between charge voltages and discharge voltages,
and 2) $R(Q) = (V_c(Q) - V_d(Q)) / (I_c(Q) - I_d(Q))$, corresponding to the status of internal resistance as $Q$ varies.
These cycle-level features serve as the basic elements for calculating intra-cell and inter-cell differences.

There are two critical processing steps to ensure the effectiveness of calculating these differences from cycle-level feature maps.
This first is to normalize the capacity of all batteries by their respective norminal capacities, ensuring the range of $Q$ falls within $[0, 1]$, to adapt to various battery types with very different nominal capacities.
The second is to eliminate the data noise introduced by abrupt changes in current and voltage signals due to the variations in cycling protocols.
We employ a rolling-median-based filter to alleviate this issue:
\begin{align}
\mathbf{M} &= \text{rolling\_median} (\mathbf{r}, w), \\
\Delta\mathbf{M} &= \text{abs}(\text{rolling\_median}(\mathbf{M}, w)), \\
 r_t^{new} &=
  \begin{cases}
   M_t  & \text{if } \Delta M_t > 3 \times \text{median}(\Delta \mathbf{M}), \\
   r_t        & \text{otherwise.}
  \end{cases}
\end{align}
Here $\mathbf{r}$ is the raw signal and $w$ represents the window size of the rolling operations.

\subsection*{Unifying Intra-cell and Inter-cell Learning}

Given the intra-cell feature of a battery cell, denoted as $\bm{x}$, and its cycle life $y$, we want to optimize the following objective to obtain a perfect cycle life predictor.
\begin{equation}
\label{eq:exp_reg_obj}
\min_{\bm{\theta}} \mathbb{E}_{(\bm{x}, y) \in D} 
    \left\lVert f_{\bm{\theta}}(\bm{x}) - y \right\rVert_2^2,
\end{equation}
where $D$ denotes the data distribution, $f_{\bm{\theta}}$ is an encoding function parameterized by ${\bm{\theta}}$, and $\mathbf{w}_f$ is the weight of the last prediction layer.
In practice, we need to perform the empirical risk minimization over limited training instances, that is
\begin{equation}
\label{eq:emp_reg_obj}
\min_{\bm{\theta}} \sum_{i=1}^N \left\lVert f_{\bm{\theta}}(\bm{x}_i) - y_i \right\rVert_2^2.
\end{equation}

However, battery lifetime prediction is a special task that suffers from the data-scarce challenge due to the necessity of non-linear modeling and the huge cost of obtaining data labels.
In this scenario, instantiating $f_{\bm{\theta}}$ as a neural network has a high risk of overfitting.
To alleviate the data-scarce issue, we consider modeling the differences between two distinct cells:
\begin{equation}
\label{eq:exp_cont_obj}
\min_{\bm{\theta}} \mathbb{E}_{(\Delta \bm{x}, \Delta y)} 
    \left\lVert g_{\bm{\phi}}(\Delta \bm{x}) - \Delta y \right\rVert_2^2,
\end{equation}
where $\Delta \bm{x} = \bm{x} - \bm{x'}$,
$\Delta y = y - y'$,
$(\bm{x}, y)$ and $(\bm{x'}, y')$ are independently sampled from $D$,
and $g_{\bm{\phi}}$ is a function parameterized by ${\bm{\phi}}$ that operates in the space of $\Delta \bm{x}$.

The inter-cell learning formulation~\eqref{eq:exp_cont_obj} assumes that the differences in feature representations for any pair of battery cells hold a unified relationship with the differences in their lifetimes.
Specifically, we can establish a clear connection between $f_{\bm{\theta}}$ and $g_{\bm{\phi}}$ in the linear setting.
For example, if $y$ is zero-centered (can be done via pre-processing), and the optimal solution for the original objective~\eqref{eq:exp_reg_obj} is $f_{{\bm{\theta}}^*} (\bm{x}) = {\bm{w}^*}^T \bm{x} $, then it is easy to verify that $g_{{\bm{\phi}}^*} = f_{{\bm{\theta}}^*}$ is also the optimal solution for the contrastive objective~\eqref{eq:exp_cont_obj} because $\mathbb{E}_{(\Delta \bm{x}, \Delta y)} \left\lVert {\bm{w}^*}^T \Delta \bm{x} - \Delta y \right\rVert_2^2$ can be decomposed into
\begin{equation}
\label{eq:obj_rel_ori_cont}
\mathbb{E}_{(\bm{x}, y) \in D} \left\lVert {\bm{w}^*}^T \bm{x} - y \right\rVert_2^2
+
\mathbb{E}_{(\bm{x}', y') \in D} \left\lVert {\bm{w}^*}^T \bm{x}' - y' \right\rVert_2^2.
\end{equation}

In the non-linear setting, such as using neural networks as function approximators, it is intractable to establish the exact connection between $f_{\bm{\theta}}$ and $g_{\bm{\phi}}$, which leads to two separate optimization procedures.
Inspired by the same optimality of the objectives~\eqref{eq:exp_reg_obj} and~\eqref{eq:exp_cont_obj} under the linear setting, we propose to share the last linear layer of $f_{\bm{\theta}}$ and $g_{\bm{\theta}}$ when using neural networks, that is
\begin{equation}
\label{eq:neural_f_g}
f_{\bm{\theta}}(\bm{x}) = \bm{w}^T h_{\bm{\theta}} (\bm{x}), \quad
g_{\bm{\phi}}(\Delta \bm{x}) = \bm{w}^T h_{\bm{\phi}} (\Delta \bm{x}),
\end{equation}
where $\bm{w}$ is the shared parameter, $h_{{\bm{\theta}}}(\cdot)$ and $h_{{\bm{\phi}}}(\cdot)$ are two separate neural networks parameterized by $\bm{\theta}$ and $\bm{\phi}$, respectively.
In this way, we can connect the optimization of the objectives~\eqref{eq:exp_reg_obj} and~\eqref{eq:exp_cont_obj} via the shared parameter $\bm{w}$ and enjoy the complementarity between the initial intra-cell learning and the new inter-cell modeling.

Moreover, we also need to perform empirical risk minimization to mimic the desired objective~\eqref{eq:exp_cont_obj}.
Specifically, given that we only have $N$ training instances, we use $N(N-1)$ pairs of instances to substitute the expectation over independently sampled instance pairs.
Together with the original regression, we have the following joint empirical risk minimization problem:
\begin{equation}
\label{eq:emp_joint_obj}
\min_{\bm{w}, {\bm{\theta}}, {\bm{\phi}}}
\sum_{i=1}^N \left\lVert \bm{w}^T h_{\bm{\theta}}(\bm{x}_i) - y_i \right\rVert_2^2
+ \lambda 
\sum_{i=1}^N \sum_{\substack{j=1 \\ {j \ne i}}}^N \left\lVert \bm{w}^T h_{\bm{\phi}}( \Delta \bm{x}_{i,j}) - \Delta y_{i,j} \right\rVert_2^2,
\end{equation}
where $\Delta \bm{x}_{i,j} = \bm{x}_i - \bm{x}_j$, $\Delta y_{i,j} = y_i - y_j$, and $\lambda$ is a hyper-parameter to balance two regression objectives.

After the optimization stage, we can leverage the neural networks for encoding intra-cell and inter-cell differences to make predictions.
Given an unseen instance $\bm{x}$, we have
\begin{equation}
\label{eq:infer}
\hat{y}^o = \bm{w}^T h_{\bm{\theta}} (\bm{x}), \quad
\hat{y}^c = \bm{w}^T h_{\bm{\phi}} (\bm{x} - \bm{x}') + y',
\end{equation}
where $\hat{y}^o$ and $\hat{y}^c$ are the predictions in the original and contrastive space, respectively, and $(\bm{x}', y')$ can be sampled from the training instances.
Last, we predict the lifetime as $\hat{y} = \alpha \hat{y}^o + (1 - \alpha) \hat{y}^c$, where $\alpha$ is a hyper-parameter to balance the two types of predictions.

\subsection*{Selection strategy for the reference cells}

\begin{figure*}[!t]
    \includegraphics[width=\linewidth]{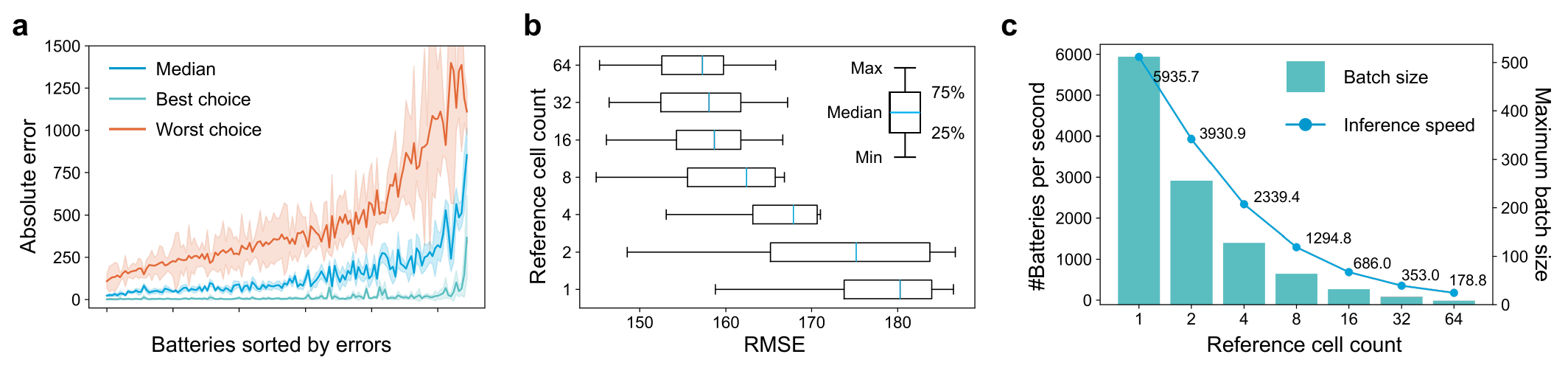}
    \caption{Selection strategy for reference cells and the impact of the number of reference cells for each battery. \textbf{a}: absolute prediction error for each battery using the optimal, worst, and median of the reference cells. The shades represents the variance among eight random seeds. \textbf{b}: a plot of prediction error with respect to different number of the reference cells. \textbf{c}: visualization of the inference speed and memory consumption with different number of reference cells.}
    \label{fig:ref_strategy}
\end{figure*}

While the reference signals of intra-cell models are restricted to their own cycles, inter-cell modeling allows flexible selection of reference batteries.
This additional freedom enables us to enhance the predictive performance through the careful choice of reference cells.

Figure \ref{fig:ref_strategy}\textbf{a} demonstrates the substantial impact of different reference cell selection strategy on the prediction error for each battery.
The gap between the least favorable and optimal reference choices can lead to an absolute error exceeding $1500$ cycles.
The rules for the optimal reference choice of reference varies among the target batteries, as shown in Figure S6.
In \methodname, we employ the median of the predictions of a reference battery group of size 32 as the final prediction, which runs at 353 target batteries per second during inference.
Using this reference ensemble strategy, both the mean and variance of the inference error decrease as the number of reference batteries increases.
Impressively, this choice leads to a significant improvement compared to the worst-case scenario, resulting in an absolute error reduction of more than $1000$ cycles for specific cells.
However, a substantial gap still exists between the median choice and the optimal one.
Closing this gap requires the model to intelligently select the most suitable reference cell for each battery.

\figurename~\ref{fig:ref_strategy}\textbf{b} illustrates the relationship between the number of reference cells and the prediction error of InterCD.
As the number of reference cells increases, both the average prediction error and the variance between different random seeds decrease.
This indicates that a larger number of reference cells simultaneously enhances the model's accuracy and robustness.

\figurename~\ref{fig:ref_strategy}\textbf{c} depicts the impact of the reference cell count on the inference speed and batch size for \methodname.
With an increase in reference size, there is a corresponding rise in inference time and a reduction in batch size.
A large number of reference cells leads to higher computational cost, where the number of batteries processed in parallel decreases from $512$ to $8$.
However, even with a very large reference size of $64$ for each cell, the inference speed of the model remains notably fast, operating at $178.8$ cells per second.
This underscores the capability of \methodname as a real-time battery lifetime predictor.

\subsection*{Network architecture design and experimental setups}

The input features $\mathbf{x}$ have dimensions of $(B, 6, H, W)$, where $B$ represents batch size, $H$ stands for cycle count, and $W$ denotes the dimension of interpolated normalized capacity.
The encoder contains two 2D convolutional layers, each followed by an average pooling layer and a ReLU activation layer.
Different from computer vision applications, our use of average pooling allows the model to focus on the overall characteristics of the difference curves, rather than abrupt changes in specific locations such as spikes caused by data misalignment.
The end of the encoder is a fully connected layer that compresses the spatial dimensions $(H, W)$ to $1$, facilitating subsequent predictions through the shared linear regression layer.

We employ a hidden layer dimension of $32$ across all models.
For all neural network models, we conduct eight runs using random seeds ranging from $0$ to $7$ to evaluate their performance under different initialization conditions.
In each experiment, $32$ reference cells are employed for inference and we take the median of these reference cells as the final prediction for the InterCD branch.
All experiments are executed on a server with 8 GTX 3090 GPUs, and we leverage PyTorch's timer to conduct repeated runtime assessment.

\subsection*{Evaluation metrics}


To comprehensively evaluate model performance, this work utilizes three standard error metrics: root mean square error (RMSE), absolute percentage error (APE), and mean absolute percentage error (MAPE).
The RMSE provides an absolute measure of the difference between predicted values ($\mathbf{y}$) and ground truth labels ($\hat{\mathbf{y}}$). It is calculated by first finding the squared error between each paired prediction and label, summing these, then taking the square root of the average squared error across all samples $N$:
\begin{equation}
\mathrm{RMSE}(\mathbf{y}, \hat{\mathbf{y}}) = \sqrt{\frac{1}{N}\sum_{n=1}^N (y_n - \hat{y}_n)^2}
\end{equation}
Lower RMSE values indicate better model performance, with a value of 0 corresponding to perfect predictions. However, RMSE lacks normalization and thus can be difficult to interpret across datasets.
APE provides a relative percentage measure of the deviation between each prediction-label pair. It is defined as the absolute ratio between the error and the ground truth label:
\begin{equation}
\mathrm{APE}(y_n, \hat{y}_n) = \left|\frac{y_n - \hat{y}_n}{\hat{y}_n}\right|
\end{equation}
Unlike RMSE, APE is unitless and normalized, making it more interpretable across tasks. However, it lacks a global view of performance across samples.
Finally, MAPE reports the mean of APE across all predictions to quantify overall performance. It is calculated by taking the average absolute percentage error:
\begin{equation}
\mathrm{MAPE}(\mathbf{y}, \hat{\mathbf{y}}) = \frac{1}{N}\sum_{n=1}^N\mathrm{APE}(y_n, \hat{y}_n)
\end{equation}
Lower MAPE indicates better generalization. Together, these metrics provide comprehensive and complementary insights into model performance.

\subsection*{Cell-level error analysis}

\begin{figure*}
    \includegraphics[width=\linewidth]{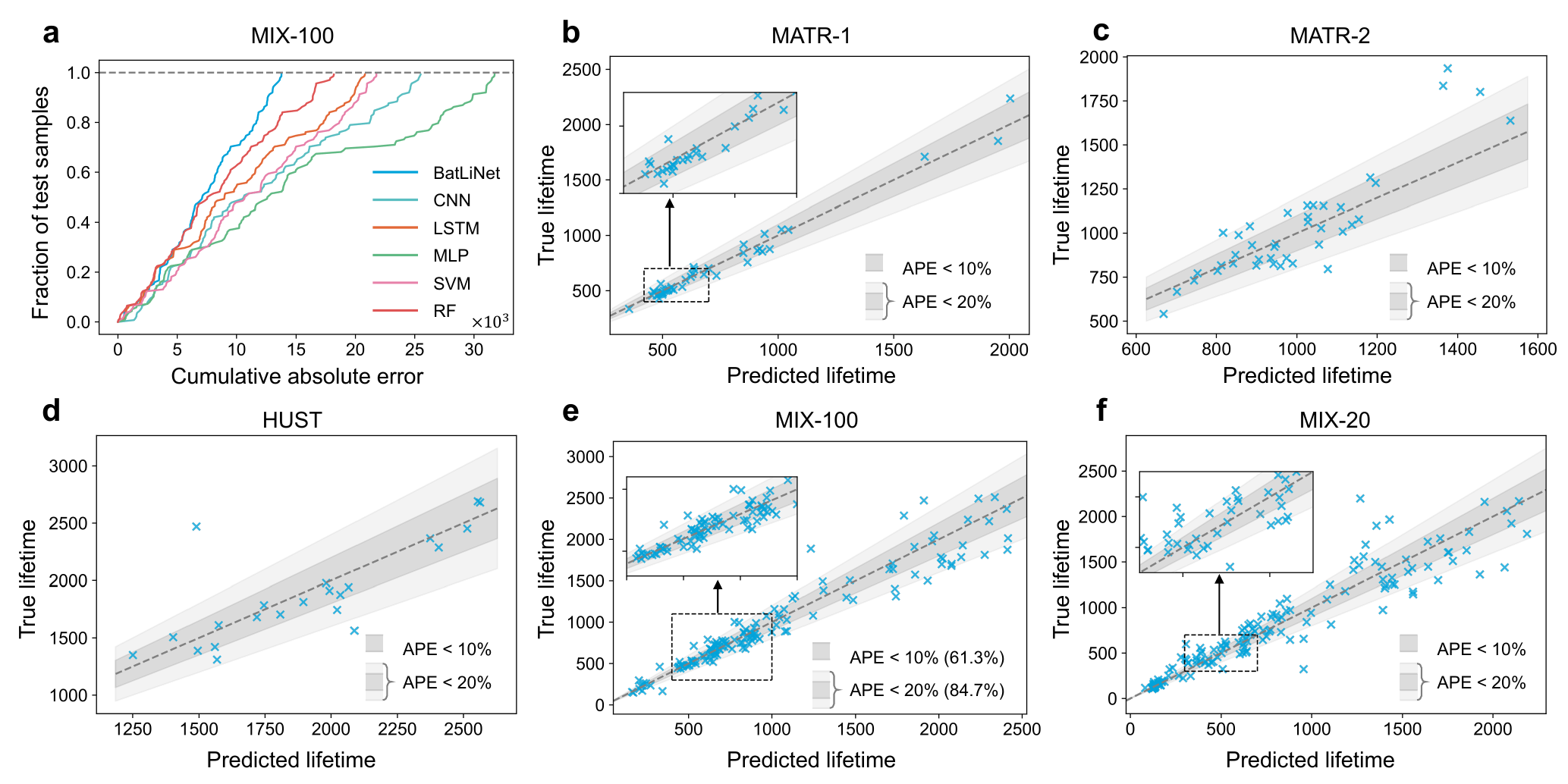}
    \caption{Cell-level error analysis. \textbf{a}: Cell-level error comparison between \methodname and baselines on \texttt{MIX-100}. \textbf{b-f}: plots of predicted lifetime compared with true lifetime on the five benchmark datasets.}
    \label{fig:instance_err}
\end{figure*}

Both RMSE and MAPE reflects the average error on the dataset scale, which can conceal anomalously large errors in individual samples.
Here we study the error distribution on the cell-level to provide a more granular view of the predictive performance of \methodname.

\figurename~\ref{fig:instance_err}\textbf{a} plots the cumulative absolute error with respect to the fraction of test samples in \texttt{MIX-100} dataset.
For each method, the test cells are added in the same order.
The error curves of the random forest (RF) and \methodname initially coincide for the first approximately $50\%$ of test samples, after which the error of RF begins to escalate at a faster rate than that of \methodname.
This suggests that RF accurately captures the patterns within a subset of the samples, whereas \methodname excels at capturing the patterns in the remaining portion.
The error curve slopes of other baselines are all smaller than \methodname, indicating that \methodname consistently outperforms these methods for each data point.


\figurename~\ref{fig:instance_err}\textbf{b}-\textbf{f} shows the predicted lifetimes by \methodname versus the ground truth on the five benchmarks. The errors on \texttt{MATR-1} are well controlled across all batteries, especially for low lifetime batteries where the errors are far below $10\%$.
On the \texttt{MATR-2} dataset, the majority of batteries have errors within $20\%$, but the model tends to underestimate high lifetime batteries due to distribution shift between training and testing.
The errors on the \texttt{HUST} dataset are mainly from severely underestimating an outlier, which can greatly inflate the average error as the number of test samples is small. 
On the \texttt{MIX-100} dataset, $84.7\%$ of batteries have prediction errors within $20\%$, with only a few outliers among the 137 test samples. This shows that \methodname can accurately capture complex battery degradation patterns across varying battery types and cycling conditions, leading to highly accurate and robust lifetime predictions in most cases.
The model still struggles on \texttt{MIX-20}, where many samples have large prediction biases, including both over and under-estimation. Accurate lifetime prediction remains a challenging problem in such information-scarce scenarios. 

\subsection*{Data collection and benchmark construction}

We collect the batteries that are applicable to lifetime prediction from existing data sources. The data batteries contain the voltage and current signals during the entire cycling test. Table S1 shows the statistics of the batteries collected from different data sources.
We collected the CS2 and CX2 commercial prismatic cells from the Center for Advanced Life Cycle Engineering (CALCE). For all cells, an identical charging profile was employed, utilizing a standard constant current/constant voltage procedure with a consistent current rate of 0.5C until the voltage reached 4.2V. Subsequently, the voltage was maintained at 4.2V until the charging current fell below 0.05A. The discharge cut-off voltage for all batteries was set at 2.7V. The data for all cells are publicly available.

The 124 commercial lithium-ion batteries of MATR-1 and MATR-2 were cycled to failure under fast-charging conditions. Specifically, these lithium-ion phosphate (LFP)/graphite cells, developed by A123 Systems (APR18650M1A), were cycled in horizontal cylindrical fixtures using a 48-channel Arbin LBT potentiostat. The experiments occurred within a temperature-controlled chamber with forced convection, set at 30°C. The cells have a nominal capacity of 1.1 Ah and a nominal voltage of 3.3 V. The cells were subjected to a two-stage charging protocol, involving a rapid charge rate until reaching 80\% capacity, followed by a charge at 1C. The upper and lower cutoff potentials were set at 3.6 V and 2.0 V, respectively, and all cells underwent discharge at 4C. The data for all cells are publicly available.

We collected additional 55 LFP/graphite cells that are used in Ref. CLO. All cells underwent an optimized six-step, 10-minute fast-charging protocol. Other specifications, including the battery model, environment temperature, and discharge protocols are the same as MATR-1 and MATR-2. The data for all cells are publicly available.

RWTH dataset comprises time-series electrical and temperature signals from a cyclic aging test involving 48 lithium-ion battery cells. During the experiment, 48 cells of identical configuration underwent aging using the same profile under uniform conditions. The cells utilized are Sanyo/Panasonic UR18650E cylindrical cells, commercially available and mass-produced through a well-established fabrication process, which is designed with a carbon anode and NMC as the cathode material. The data for all cells are publicly available.

UL-PUR dataset consists of 10 commercial 18650 cells with a graphite negative electrode and an NCA positive electrode. The cells were cycled at 0.5C at 2.7-4.2V (0-100\% SOC) or 2.5-96.5\% SOC at room temperature, until experiencing a 20\% capacity fade. The HNEI dataset comprises 14 commercially available 18650 cells featuring a graphite negative electrode and a composite positive electrode made up of NMC and LCO. These cells were cycled at 1.5C to achieve a depth of discharge of 100\% for over 1000 cycles at room temperature. The SNL dataset consists of 61 commercial 18650 NCA, NMC, and LFP cells cycled to 80\% capacity. These cells experience different temperature, depth of discharge, and discharge current on the long-term degradation process. These three datasets currently are available at Battery Archive upon request.

Table \ref{tab:raw_data_stats} and \ref{tab:benchmark_stats} shows the statistics of the data sources and our organized datasets. The code for organizing and preprocessing the datasets are publicly available.

\begin{table*}
\caption{Statistics of data sources used in this study.}
\label{tab:raw_data_stats}
\begin{tabular}{@{}lcccccc@{}}
\toprule
                & \textbf{\begin{tabular}[c]{@{}l@{}}Cell\\ count\end{tabular}} & \textbf{\begin{tabular}[c]{@{}l@{}}Electrode\\ materials\end{tabular}} & \textbf{\begin{tabular}[c]{@{}l@{}}Charge\\ protocols\end{tabular}} & \textbf{\begin{tabular}[c]{@{}l@{}}Discharge\\ protocols\end{tabular}} & \textbf{\begin{tabular}[c]{@{}l@{}}Environment\\ temperatures\end{tabular}} & \textbf{\begin{tabular}[c]{@{}l@{}}Package\\ structures\end{tabular}} \\ \midrule
\textbf{CALCE}           & 13                                                            & 1                                                                      & 1                                                                   & 2                                                                      & 1                                                                           & 1                                                                     \\
\textbf{HNEI}   & 14                                                            & 1                                                                      & 1                                                                   & 1                                                                      & 1                                                                           & 1                                                                     \\
\textbf{HUST}   & 77                                                            & 1                                                                      & 1                                                                   & 77                                                                     & 1                                                                           & 1                                                                     \\
\textbf{MATR-1} & 84                                                            & 1                                                                      & 61                                                                  & 1                                                                      & 1                                                                           & 1                                                                     \\
\textbf{MATR-2} & 81                                                            & 1                                                                      & 47                                                                  & 1                                                                      & 1                                                                           & 1                                                                     \\
\textbf{CLO}    & 45                                                            & 1                                                                      & 9                                                                   & 1                                                                      & 1                                                                           & 1                                                                     \\
\textbf{RWTH}   & 48                                                            & 1                                                                      & 1                                                                   & 1                                                                      & 1                                                                           & 1                                                                     \\
\textbf{SNL}    & 61                                                            & 3                                                                      & 1                                                                   & 4                                                                      & 3                                                                           & 1                                                                     \\
\textbf{UL-PUR} & 10                                                            & 1                                                                      & 1                                                                   & 1                                                                      & 1                                                                           & 1                                                                     \\ \midrule
\textbf{Total}  & 401                                                           & 5                                                                      & 83                                                                  & 85                                                                     & 5                                                                           & 2                                                                     \\ \bottomrule
\end{tabular}
\end{table*}

\begin{table*}
\caption{Statistics of the organized datasets used in this study. Note that we filtered out batteries that is not reached their end of life during the early cycles.}
\label{tab:benchmark_stats}
\begin{tabular}{@{}lccccccc@{}}
\toprule
                 & \textbf{\begin{tabular}[c]{@{}l@{}}Cell\\ count\end{tabular}} & \textbf{\begin{tabular}[c]{@{}l@{}}Train cell\\ count\end{tabular}} & \textbf{\begin{tabular}[c]{@{}l@{}}Test cell\\ count\end{tabular}} & \textbf{\begin{tabular}[c]{@{}l@{}}End of life\\ percentage\end{tabular}} & \textbf{\begin{tabular}[c]{@{}l@{}}Early cycle\\ count\end{tabular}} & \textbf{\begin{tabular}[c]{@{}l@{}}Max\\ lifetime\end{tabular}} & \textbf{\begin{tabular}[c]{@{}l@{}}Min\\ lifetime\end{tabular}} \\ \midrule
\textbf{MATR-1}  & 83                                                            & 41                                                                  & 42                                                                 & 80\%                                                                      & 100                                                                  & 2237                                                            & 300                                                             \\
\textbf{MATR-2}  & 81                                                            & 41                                                                  & 40                                                                 & 80\%                                                                      & 100                                                                  & 2160                                                            & 300                                                             \\
\textbf{HUST}    & 77                                                            & 55                                                                  & 22                                                                 & 80\%                                                                      & 100                                                                  & 2691                                                            & 1140                                                            \\
\textbf{MIX-100} & 342                                                           & 205                                                                 & 137                                                                & 80\%                                                                      & 100                                                                  & 2691                                                            & 148                                                             \\
\textbf{MIX-20}  & 354                                                           & 207                                                                 & 147                                                                & 90\%                                                                      & 20                                                                   & 2323                                                            & 104                                                             \\ \bottomrule
\end{tabular}
\end{table*}

\end{document}